\documentstyle[epsfig]{elsart}
\begin{document}
\begin{frontmatter}

\begin{flushright}
L3 Note 2135
\end{flushright}

\title{Radiative Tau Lepton Pair Production 
as a Probe of Anomalous Electromagnetic 
Couplings of the Tau}

\author{S. S. Gau},
\author{T. Paul\thanksref{tom_mail}},
\author{J. Swain\thanksref{john_mail}},
\author{L. Taylor\thanksref{lucas_mail}}
\address{Department of Physics, Northeastern University, 
Boston, MA 02115, USA}
\thanks[tom_mail]{E-mail: tom.paul@cern.ch}
\thanks[john_mail]{E-mail: lucas.taylor@cern.ch}
\thanks[lucas_mail]{E-mail: swain@neuhp1.physics.neu.edu}
%
%
%
\begin{abstract}                    
We calculate the squared matrix element for the process
$e^+e^-\rightarrow\tau^+\tau^-\gamma$ allowing for anomalous
magnetic and electric dipole moments at the $\tau\tau\gamma$ vertex.
No interferences are
neglected and no approximations of light fermion masses are made. 
We show that anomalous moments affect not only the cross section,
but also the shape of the photon energy and angular distributions. 
We also demonstrate that in the case of the anomalous magnetic
dipole moment, the contribution from interference involving
Standard Model and anomalous amplitudes is significant compared to
the contribution from anomalous amplitudes alone.  A program to
perform the calculation is available and it may be employed as a
Monte Carlo generator.  %
\end{abstract}

\end{frontmatter}
%
%
\section{Introduction}

In the Standard Model, the charged leptons are identical in all
respects except for their masses and their distinct and conserved
lepton numbers.  There is, however, no experimentally verified
explanation for why there are three generations of leptons nor why
their masses are so different.  New insight might be forthcoming if
the leptons were observed to have substructure, which could be
manifest as an anomalous magnetic or electric dipole moment. The
anomalous moments for the electron and muon are already measured to
be in extremely good agreement with the predictions of
QED\cite{COHEN87A,ABDULLAH90A,BAILEY78A}. Anomalous moments of the
tau are relatively poorly measured and are of particular interest
since, as the heaviest lepton, the tau may exhibit the most
sensitivity to new physics.

In general a photon may couple to a tau through its electric
charge, magnetic dipole moment, or electric dipole moment
(neglecting possible anapole moments).  We may parametrise this
coupling with the following matrix element:
\begin{equation}
\langle \tau(p^\prime) | J_\mu | \tau(p) \rangle =  
\bar{u}(p^\prime) \Gamma_\mu u(p). 
\label{equ:matrixel} 
\end{equation}
$\Gamma_\mu$ represents the most general Lorentz-invariant form
of the coupling 
of a tau to a photon:
\begin{equation}
\Gamma_\mu =          F_1 (q^2)           \gamma_\mu 
           + i \frac{ F_2 (q^2)}{2m_\tau} \sigma_{\mu\nu} q^\nu 
           -          F_3 (q^2)           \sigma_{\mu\nu} q^\nu \gamma_5,
\label{equ:formfactors}
\end{equation}

where $m_\tau$ is the mass of the $\tau$ lepton, and $q = p^\prime - p$ 
is the momentum transfer. As can be verified using the
Gordon decomposition, the $q^2$-dependent form-factors, $F_i(q^2)$,
have familiar interpretations for $q^2=0$ and with the $\tau$ on
mass-shell: $F_1(0) \equiv q_\tau$ is the electric charge of the
tau, $F_2(0) \equiv a_\tau=(g-2)/2$ is the static
anomalous magnetic moment of the tau (where $g$ is
the gyromagnetic ratio), and $F_3(0) \equiv
d_\tau/q_\tau$, where $d_\tau$ is the static electric dipole moment
of the tau and $q_\tau$ is its charge.  Hermiticity of the
electromagnetic current forces all the $F_i$ to be real.

In the Standard Model, $a_\tau$ is non-zero due to loop effects
and  has been calculated to be\cite{SAMUEL91A} 
\begin{equation} 
 a_\tau = \alpha/2\pi + \ldots = 0.001\,177\,3(3).
\end{equation} 
A non-zero value of $d_\tau$ is forbidden by both $P$ invariance
and $T$ invariance. Assuming $CPT$ invariance, observation of a
non-zero value of $d_\tau$ would  imply $CP$ violation in the
$\tau$ system.

Several limits on $F_2(q^2)$ and $F_3(q^2)$ at various values of
$q^2$ have been reported
\cite{SILVERMAN83A,ESCRIBANO97A,ESCRIBANO94A,ESCRIBANO93A,GRIFOLS91A,DELAGUIA90A}.  
The $q^2$ dependence of these form factors depends strongly on the 
hypothetical mechanism giving rise to anomalous values of $F_2$ and
$F_3$. It is commonly assumed that $F_2(0)$ corresponds to the
static anomalous magnetic moment and $F_3(0)$ the static electric
dipole moment, since the radiated photon has zero $q^2$. This
condition is necessary but not sufficient; the tau must also be
on-shell on both sides of the $\tau\tau\gamma$ vertex which is only
valid in the limit of vanishing photon energy.  This condition is
satisfied for the precession measurements of electrons and
muons~\cite{COHEN87A}. None of the measurements of tau anomalous 
moments published thus far satisfies both of these conditions, so
we  urge caution in the interpretation of the limits obtained on
the form factors $F_2$ and $F_3$ in terms of the static tau 
properties, $a_\tau$ and $d_\tau$.

Table~\ref{tab:dms} shows published theoretical predictions and experimental 
limits on $F_2$ and $F_3$ for electrons, muons and taus.  For the case of
taus, the experimental limits of reference~\cite{SILVERMAN83A} are derived
from the $\mathrm{e^+e^-}\rightarrow\tau^+\tau^-$ total cross section at PETRA 
at $q^2$ up to $(37\mathrm{GeV})^2$, those of reference~\cite{ESCRIBANO97A}
are calculated from $\Gamma(Z \rightarrow \tau^+\tau^-)$, and those of~\cite{GRIFOLS91A}
are found using the total $\mathrm{e^+e^-}\rightarrow\tau^+\tau^-\gamma$
cross section from early LEP data.  We note that this last result
is the only one which corresponds to a direct measurement at 
$q^2=0$.

This paper describes a tree-level calculation of the process 
${\mathrm{e^+e^-}}\rightarrow\tau^+\tau^-\gamma$ which accounts for
the  effects of anomalous magnetic and electric dipole couplings.
We present total cross sections as well as energy and angular
distributions of the radiated photon which will
be useful for measuring $F_2(0)$ and $F_3(0)$.
We also demonstrate the importance of including
interference  involving Standard Model and anomalous amplitudes,
which has not been considered experimentally up until now.  The
calculation described here may be employed as an event generator,
making it possible to simulate detector effects and properly
account for acceptance, as is necessary for a meaningful
interpretation of the data.

\begin{table}[hbt!]
\setlength{\tabcolsep}{0.5mm}
\renewcommand{\arraystretch}{1.6}
\begin{center}
{%
\begin{tabular}{|c|lrlrlclc|}  \hline 
%
                 & ~~Theory:~~ & &  & $F_2$   &    $(0)$             & $=$ & $0.001\,159\,652\,46(15)$                         & \cite{KINOSHITA81A} \\  \cline{2-9}
~~e~~            & ~~Expt:~~   & &  & $F_2$   &    $(0)$             & $=$ & $0.001\,159\,652\,193(10)$                        & \cite{COHEN87A}     \\  \cline{2-9}
                 & ~~Expt:~~   & &  & $eF_3$  &    $(0)$             & $=$ & $(-2.7 \pm 8.3) \cdot 10^{-27}\,e\,{\mathrm{cm}}$ & \cite{ABDULLAH90A}  \\  \hline\hline
                 & ~~Theory:~~ & &  & $F_2$   &    $(0)$             & $=$ & $0.001\,165\,920\,2(20)$                          & \cite{HUGHES85A}    \\  \cline{2-9}
~~$\mu$~~        & ~~Expt:~~   & &  & $F_2$   &    $(0)$             & $=$ & $0.001\,165\,923\,0(84)$                          & \cite{COHEN87A}     \\  \cline{2-9}
                 & ~~Expt:~~   & &  & $eF_3$  &    $(0)$             & $=$ & $( 3.7 \pm 3.4) \cdot 10^{-19}\,e\,{\mathrm{cm}}$ & \cite{BAILEY78A}    \\  \hline\hline
                 & ~~Theory:~~ & &  & $F_2$   &    $(0)$             & $=$ & $0.001\,177\,3(3)$                                & \cite{SAMUEL91A}    \\  \cline{2-9}
~~$\tau$~~       & ~~Expt:~~   & &  & $F_2$   &    $(q^2 \neq 0)$    & $<$ & $0.02$                                            & \cite{SILVERMAN83A} \\
                 &             &$-0.004$ & $<$ & $F_2$   &    $(q^2 \neq 0)$    & $<$ & $0.006$                                            & \cite{ESCRIBANO97A} \\
                 &             & &  & $F_2$   &    $(0)$    & $<$ & $0.11$                                            & \cite{GRIFOLS91A}   \\  \cline{2-9}
%
                 & ~~Expt:~~   & &  & $eF_3$  &    $(q^2 \neq 0)$    & $<$ & $1.4           \cdot 10^{-16}\,e\,{\mathrm{cm}}$  & \cite{DELAGUIA90A}  \\
                 &             & &  & $eF_3$  &    $(q^2 \neq 0)$    & $<$ & $0.11           \cdot 10^{-16}\,e\,{\mathrm{cm}}$  & \cite{ESCRIBANO97A} \\
                 &             & &  & $eF_3$  &    $(0)$             & $<$ & $6.0           \cdot 10^{-16}\,e\,{\mathrm{cm}}$  & \cite{GRIFOLS91A}   \\ \hline
%
%
%
%
%
%
\end{tabular}
}%
\caption{Theoretical predictions and experimental measurements of the
         anomalous magnetic moments and the experimental 
         measurements of the electric dipole moments of charged leptons.}
\label{tab:dms}
\end{center}
\end{table}


\section{Calculation of the Effects of Anomalous Couplings 
in $e^+e^-\rightarrow\tau^+\tau^-\gamma$ 
Events}~\label{sec:calculation}

\begin{figure}[htb]
\begin{center}
\epsfig{file=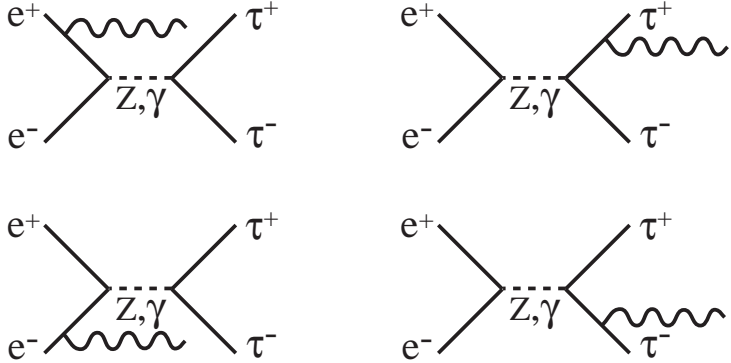,width=0.7\textwidth,clip=}
\vfill
\caption{\label{fig:feynmann}Diagrams contributing to $e^+ e^- 
\rightarrow \tau^+ \tau^- \gamma$.}
\end{center}
\end{figure}

Using the parametrisation of the electromagnetic current given  in
Eq.~\ref{equ:formfactors},  we consider all the Standard Model and
anomalous  amplitudes for the diagrams shown in
Fig.~\ref{fig:feynmann}.  The corresponding matrix element is then
evaluated using the symbolic  manipulation package FORM~\cite{FORM}
without making any simplifying assumptions. In particular, no
interference terms are neglected and no fermion masses are assumed
to be zero.  The inclusion of a non-zero tau mass is essential, as
the (significant) interference terms between the Standard Model and
the anomalous  amplitudes vanish in the limit of vanishing tau
mass. Standard Model radiative corrections are incorporated by
using the  improved Born approximation~\cite{IMPROVED_BORN}.

We use the FOWL program~\cite{FOWL} to generate points in 3-body
phase space, and for each point calculate the full squared matrix
element. This procedure yields a sample of events with
probabilistic weights which may used directly or applied in a Monte
Carlo rejection method to produce a sample of events with weights
of unity. 

Infrared divergences for collinear or low energy photons may be
naturally avoided by applying cuts on minimum photon energy and/or
opening angle between the photon and tau.  This method eliminates
potential concerns about the consistency of cancelling divergences
against vertex corrections which are calculated in QED assuming
that the anomalous couplings vanish. Moreover, this procedure
matches well the experimental reality, since  electromagnetic
calorimeters have a minimum energy cutoff and isolation cuts are
required to distinguish tau decay products from radiated photons.
%
%
\section{Anomalous Contribution to the Total Cross Section}
%
%

\begin{figure}[htb]
\begin{center}
\epsfig{file=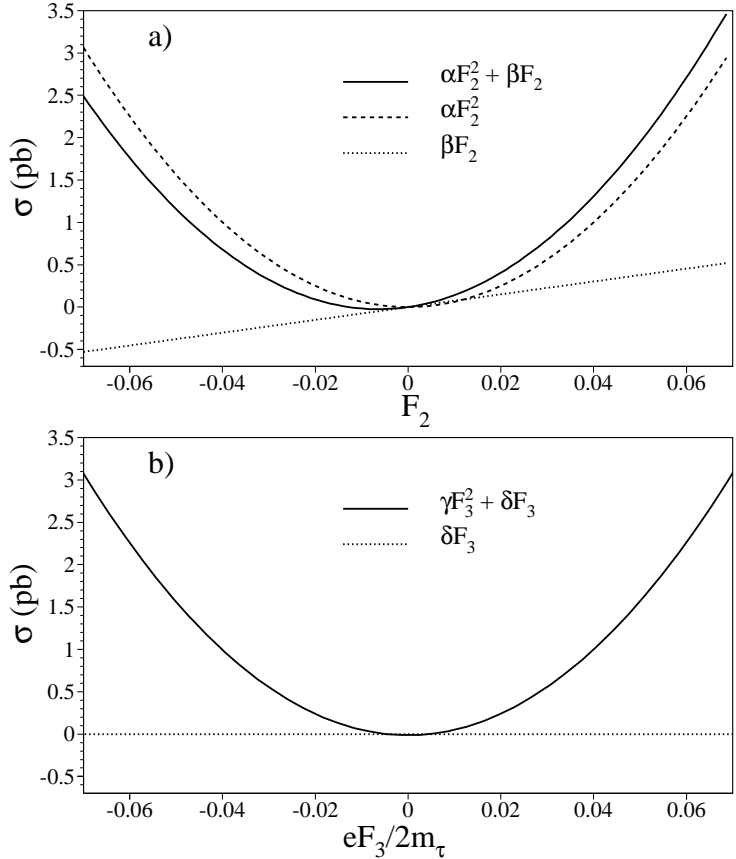,width=.8\textwidth,clip=}
\vfill
\caption{\label{fig:x_edm_mdm}Anomalous contribution to the
$e^+e^-\rightarrow\tau^+\tau^-\gamma$  cross section as a function
of a) $F_2(0)$ and b) $F_3(0)$.  The contributions from the linear
and quadratic terms are shown separately.}
\end{center}
\end{figure}

It is illustrative to isolate the anomalous contribution to the
cross section  from the purely Standard Model contribution.  The
anomalous contribution  is given by those terms in the matrix
element which contain a factor of $F_2(0)$ or $F_3(0)$:
\begin{equation} \sigma_{\mathrm{ano}} = \alpha F_2^2(0) + \beta
F_2(0) + \gamma F_3^2(0) + \delta F_3(0). 
\end{equation} 
Terms linear in $F_2(0)$ and $F_3(0)$ arise from interference between 
Standard Model and anomalous amplitudes, whereas the quadratic
terms are purely anomalous. Figure~\ref{fig:x_edm_mdm}a shows the
contribution to the total cross section~\footnote{The calculation
assumes $M_{\mathrm{Z}}$ = 91.181\,GeV, $M_\tau$ = 1.777\,GeV, and
$\sin^2(\theta_W^{eff}) = 0.2319$.} arising from these terms, with
the linear and quadratic components shown separately.  
Interestingly, the interference terms do not vanish, and in fact
are significant compared to the total  anomalous cross section for
small values of $F_2(0)$. 
%
Figure~\ref{fig:x_edm_mdm}b shows 
the anomalous cross section as a function of $F_3(0)$, with the linear term again shown separately.
Evidently the interference does vanish in the case of an electric dipole moment.  
%
%
\section{Anomalous Contribution to the Differential Cross Section}
\begin{figure}[thb]
\begin{center}
\epsfig{file=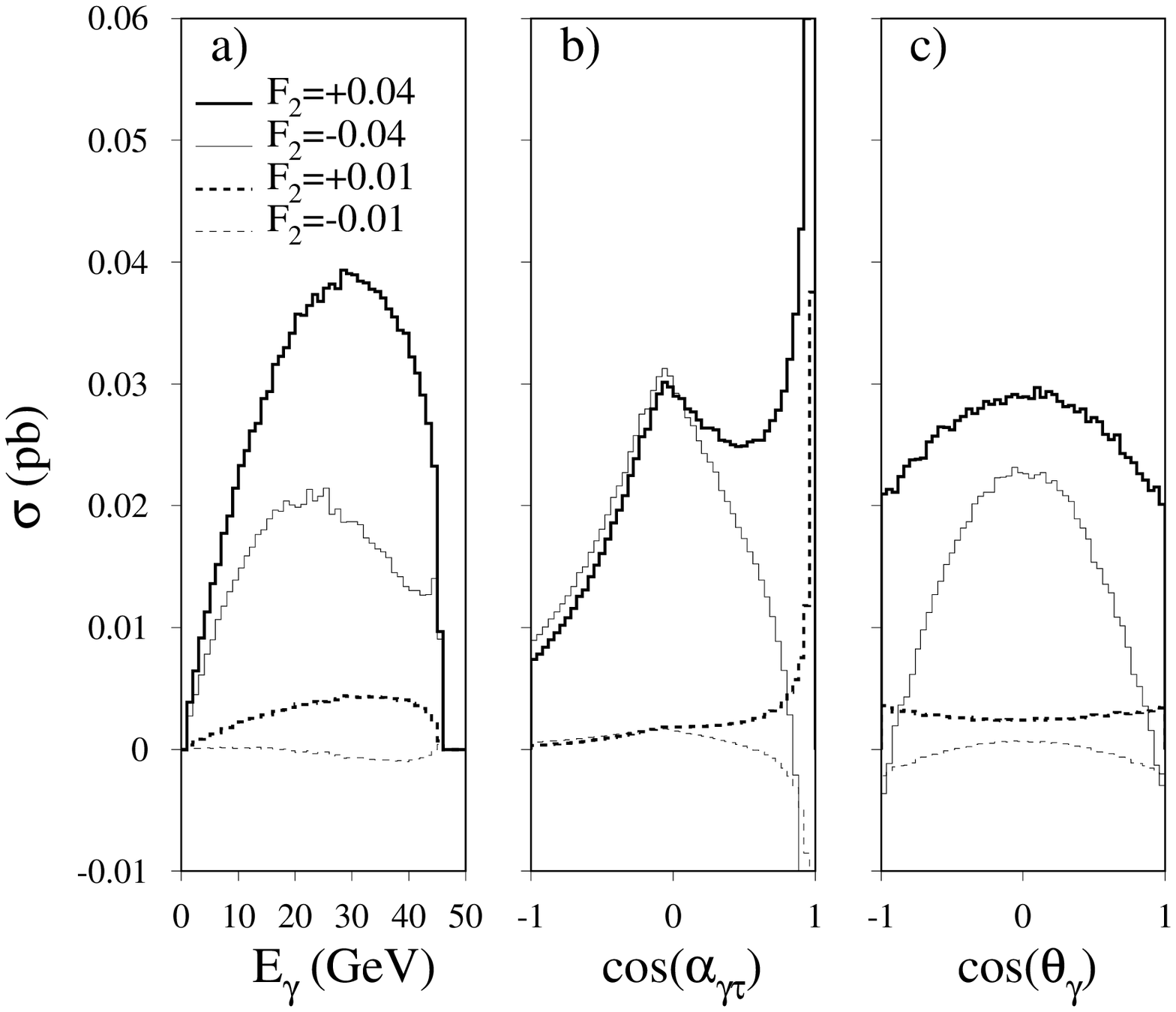,width=0.8\textwidth,clip=}
\vfill
\caption{\label{fig:e_a_t}Anomalous contribution to a) the photon
energy spectrum for various values of $F_2(0)$, b) the distribution
of the opening angle between the photon and the nearest tau, and c)
the distribution of photon polar angle.}
\end{center}
\end{figure}

\begin{figure}[thb]
\begin{center}
\epsfig{file=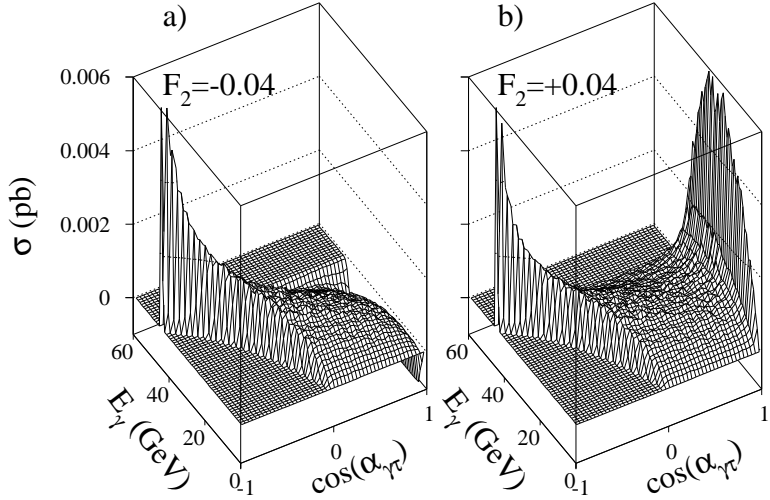,width=0.8\textwidth,clip=}
\vfill
\caption{\label{fig:e_vs_a}Anomalous contribution to the cross
section as a function of the photon energy versus the cosine of the
opening angle between the photon and tau ($\alpha_{\gamma \tau}$) 
for a) $F_2(0)=-0.04$ and b) $F_2(0)=+0.04$.}
\end{center}
\end{figure}

In addition to affecting the total cross section, anomalous
couplings also affect the shapes of energy and angular
distributions of the final state particles. Because of their
experimental accessibility, we choose to study the photon energy,
the three-dimensional opening angle between the photon and the
closest tau, and the photon polar angle with respect to the
electron direction. Figure~\ref{fig:e_a_t}a shows the anomalous
contribution to the photon  energy spectrum for various values\footnote{
Values of $F_2(0)$ used in plots in this paper were chosen
to be in the neighborhood of the expected LEP sensitivity to 
$F_2(0)$.  See section~\ref{sec:experiment}.} of $F_2(0)$.
Because of interference,  both the cross section and
spectrum shape are asymmetric around $F_2(0)=0$.
Figure~\ref{fig:e_a_t}b shows the distribution of the opening angle
between the photon and the closest tau for for various $F_2(0)$. 
The difference between spectrum shapes for $+\mid F_2(0)\mid$ and
$-\mid F_2(0)\mid$ is more striking for this distribution than for
Figure~\ref{fig:e_vs_a}, the energy and opening angle are plotted
against  one another for two different values of $F_2(0)$, showing
that these two  quantities provide independent information about
the anomalous moment. 
%
%
\section{Comparisons with Other Results}

\begin{figure}[thb]
\begin{center}
\epsfig{file=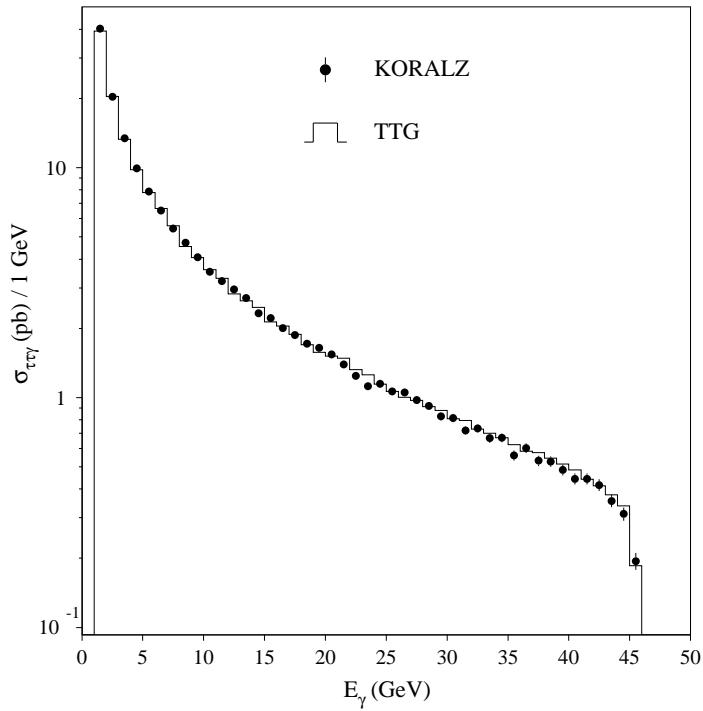,width=0.75\textwidth,clip=}
\vfill
\caption{\mbox{~}\label{fig:kz_ttg}Photon energy spectrum for 
KORALZ (dots) and TTG (histogram) for the cuts described in 
the text.}
\end{center}
\end{figure}

As a technical crosscheck, we create a sample of events weighted
using our matrix  element calculation, henceforth referred to as
``TTG,'' with $F_2(0) = F_3(0) = 0$ and  compare the total cross
section and energy and angular distributions with the predictions
of the KORALZ~\cite{KORALZ} Monte Carlo.  Since TTG provides an
${\cal O}(\alpha)$ calculation, KORALZ is also used in a mode in
which only single photon radiation is considered.  In order to
prevent infrared divergences in TTG, cuts are placed  on the photon
energy and opening angles, as discussed in
section~\ref{sec:calculation}. In particular, we require the photon
energy to be greater than 1~GeV, the cosine of the  opening angle
between the photon and the nearest $\tau$ to be less than 0.995,
and the  cosine of the polar angle of the photon with respect to
the electron direction to be less than 0.9.  The total cross
sections calculated by KORALZ and TTG for these cuts agree  to
better than $0.5\%$.  Figure~\ref{fig:kz_ttg} shows a comparison of
the photon energy  spectra computed by the two programs. 

Following an initial presentation of the results in this
paper~\cite{SWAIN96A}, an analytical calculation for the process
$e^+e^-\rightarrow\tau^+\tau^-\gamma$ was carried
out~\cite{BIEBEL96A}. This calculation makes some approximations,
but importantly  does not assume zero tau mass and does not neglect
interference between Standard Model and anomalous final states.  As
a mutual cross check, we compare our results with the total cross
section and photon energy spectrum predicted by this calculation.

\begin{figure}[thb]
\begin{center}
\epsfig{file=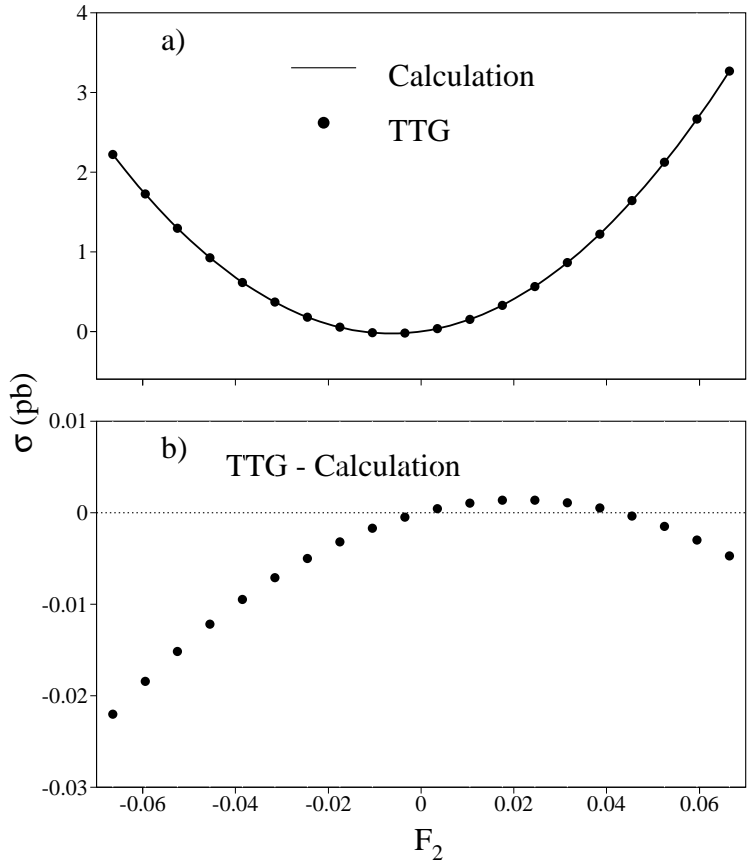,width=0.8\textwidth,clip=}
\vfill
\caption{\mbox{~}\label{fig:ttg_vs_riemann_e}a) Comparison of
anomalous cross section computed using a simplified analytical
calculation (curve) and that computed using TTG with the same
simplifications (dots). b) The difference between the two cross
sections described in a).}
\end{center}
\end{figure}

\begin{figure}[thb]
\begin{center}
\epsfig{file=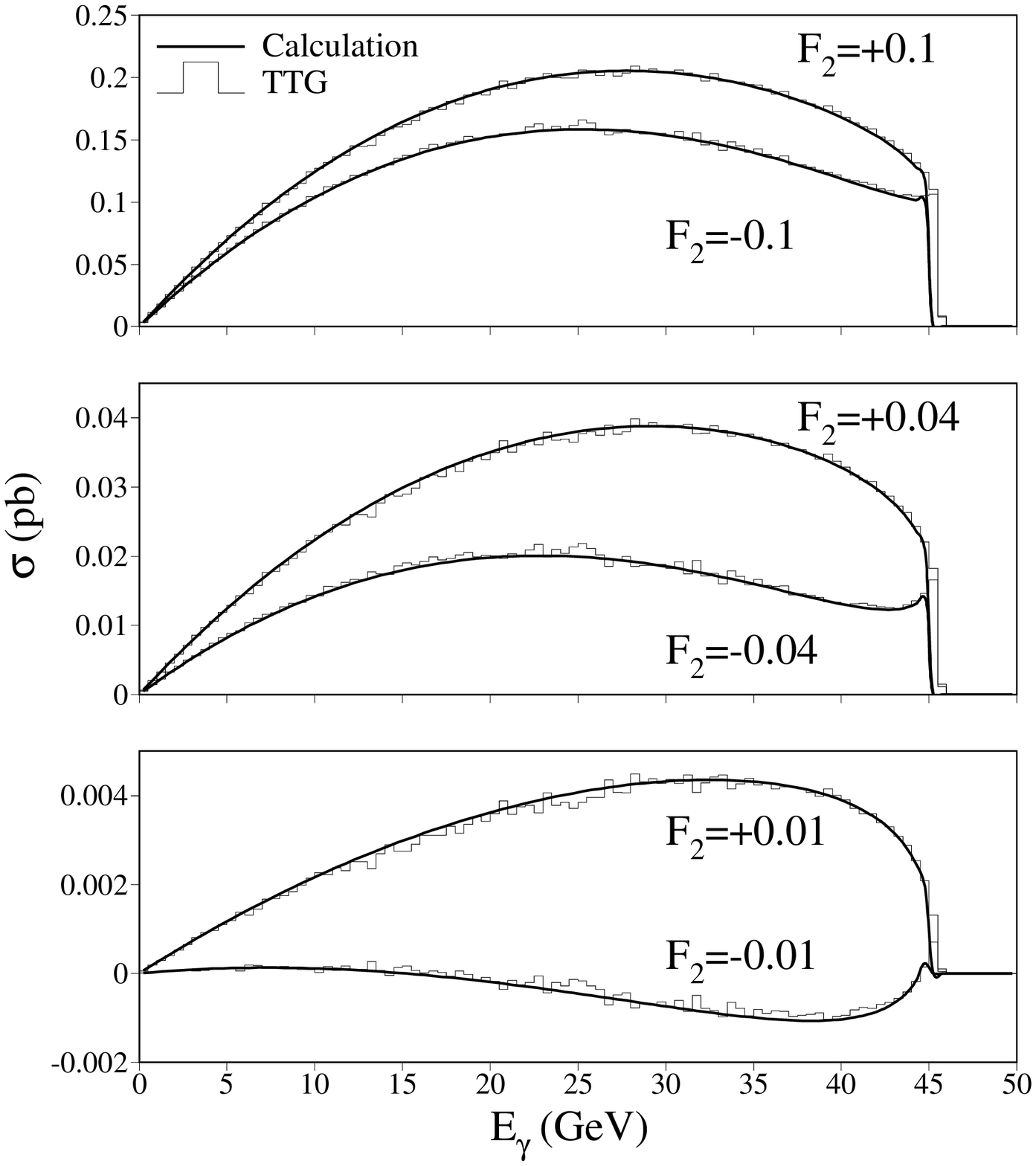,width=.8\textwidth,clip=}
\vfill
\caption{\mbox{~}\label{fig:ttg_vs_riemann_e2}Comparison for a
range of $F_2(0)$ of the photon energy spectrum  computed using a 
simplified analytical calculation (curve) and that computed using
TTG with the same simplifications (histogram).}
\end{center}
\end{figure}

The analytical calculation neglects anomalous contributions from 
initial-final state interference, from $\gamma Z$ interference, and
from $\gamma$ exchange, so for purposes of comparison we remove
these terms from our calculations as well. 
Figure~\ref{fig:ttg_vs_riemann_e} shows 
a comparison of anomalous cross sections as a function of $F_2(0)$
as determined by our simplified version of TTG and the analytical
calculation, indicating agreement at the 1\% level. 
Figure~\ref{fig:ttg_vs_riemann_e2} gives a  comparison of the
photon energy spectra for the two calculations, showing good
agreement between the spectrum shapes over a large range of
$F_2(0)$ and  the full range of photon energies accessible near the
Z resonance.  

\begin{figure}[thb]
\begin{center}
\epsfig{file=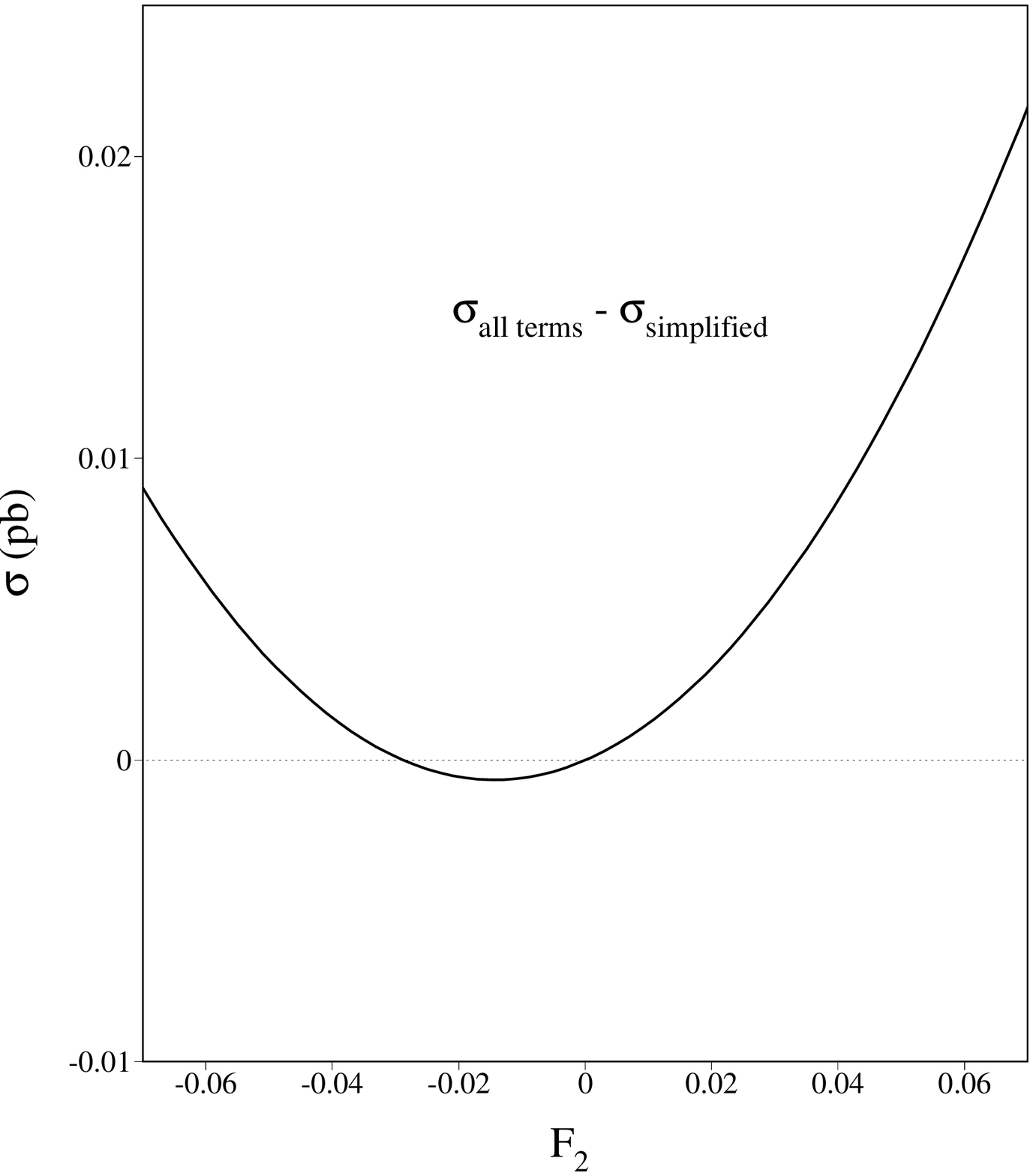,width=.7\textwidth,clip=}
\vfill
\caption{\mbox{~}\label{fig:all_vs_simple}a) Difference as a
function of $F_2(0)$ between the cross section including all 
anomalous contributions and cross section including all but
anomalous contributions from initial-final state interference,
$\gamma Z$ interference, and $\gamma$ exchange.}
\end{center}
\end{figure}

At this point, one may ask whether all the possible anomalous
contributions are  important, or whether one need only consider the
simplified case  just discussed.  To check this, we use our
calculation to determine  the full anomalous contribution to the
cross section, including all the terms  neglected in the analytical
calculation, and compare with the simplified result.
Figure~\ref{fig:all_vs_simple} shows the difference between the
cross sections as a function of $F_2(0)$ for the two cases.  The
discrepancy is roughly 1\% of the total anomalous cross section,
showing that one can safely neglect anomalous contributions from
initial-final state interference, $\gamma Z$ interference,  and
$\gamma$ exchange.   
%

\section{Experimental Considerations} \label{sec:experiment}

Previous experimental limits on $F_2(0)$ and
$F_3(0)$~\cite{GRIFOLS91A,TAU96_L3} have assumed that the
interference between the Standard Model and anomalous 
contributions to the cross section can be neglected.  As we have
shown in  Figure~\ref{fig:x_edm_mdm}, this  is not generally true
for the anomalous magnetic dipole moment. If we take $\mid F_2(0)
\mid = 0.04$, for example, then failure to account for interference
leads to an  underestimate of the anomalous contribution to the
total cross section by about 25\% if $F_2(0)$  is positive, and an
overestimate of almost 50\% if $F_2(0)$ is negative. These
discrepancies become even more pronounced for smaller $\mid F_2(0)
\mid$. 

We have shown that the photon energy, the opening angle between
the  photon and the nearest tau, and the photon polar angle each
contains independent information about $F_2(0)$ and $F_3(0)$,
including the sign of $F_2(0)$  (Figures~\ref{fig:e_a_t}
and~\ref{fig:e_vs_a}).  One should  note, however, that the
distributions presented in these figures include no cuts on photon
energies or angles.  In practice, detector geometry imposes a cut
on the photon polar angle with respect to the electron direction,
and further cuts must be applied on the photon energy and minimum
opening angle between the photon and tau in order to suppress
background from tau decay products. In particular, the unavoidable
cut  on the opening angle must be placed in exactly the region
where sensitivity to the  sign of the anomalous coupling is
highest.

Other theoretical treatments~\cite{GRIFOLS91A,BIEBEL96A} have 
provided cross sections or differential distributions, which are 
generally not sufficient for meaningful analysis of the data.   For
example, we have just seen that  the anomalous contribution to the
cross section depends strongly on the  various experimental cuts
applied, an effect which is difficult if not impossible to assess
without a full Monte Carlo simulation. The calculation described
here provides just such a tool.  It may be employed to yield
samples of 4-vectors for the final state particles in
$e^+e^-\rightarrow\tau^+\tau^-\gamma$, which can be passed to the
TAUOLA~\cite{TAUOLA} program to perform the tau decays. The
resulting final state particles can then be passed through the
detector  simulation program, which will account for such effects
as  energy loss and interactions in the detector, geometrical
acceptance, and energy and spatial resolutions, together with
obscure correlations which are  not easily handled on a statistical
basis.   Alternatively, the matrix element calculation may be used
to compute event-by-event weights for a given value of $F_2(0)$ or
$F_3(0)$ using 4-vectors from previously simulated
$e^+e^-\rightarrow\tau^+\tau^-\gamma$ events. This makes it
possible to reweight existing samples for which detector  effects
have already been simulated, thus saving considerable computing
time.

In order to estimate the sensitivity to anomalous moments expected
for the  LEP experiments running at $\sqrt{s} \approx M_Z$, we have
performed a study of simulated $\tau^+\tau^-\gamma$ events as they
would appear in the L3 detector~\cite{L3_GEANT}.  The simulated
sample corresponds to  roughly the integrated luminosity collected
by L3 at the Z peak  (about 100~$pb^{-1}$) and an anomalous
magnetic moment $F_2(0)=0$ and electric dipole moment $F_3(0)=0$.
We estimate $40\%$ efficiency for selecting $\tau^+\tau^-\gamma$
events, which is dominated by the requirement that at least one jet
be in the barrel region $(\cos \theta < 0.7)$ of the detector. 
Events  with good quality photons are selected and background from
tau decay products is rejected by asking that the transverse energy
of the photon to the nearest jet be larger than 2~GeV.  We then
perform a binned maximum likelihood fit of the selected photon
energy spectrum to an independent collection of simulated samples
which have been weighted, using the method outlined above, such
that they correspond to many different values  of $F_2(0)$.  The
fit results show that if the true value of $F_2(0)=0$, one can
expect to set a $95\%$ confidence level limit of roughly $| F_2(0)
| < 0.05$.  If $F_2(0) \neq 0$ or $F_3(0) \neq 0$,  the sensitivity
improves.  Of course, the exact value of the limit will depend on
the details of the experiment, selection applied, and the
variables  used in the fit.

\section{Conclusions}

We have carried out a calculation of the squared matrix element for
the  process $e^+e^-\rightarrow\tau^+\tau^-\gamma$, allowing for
possible anomalous magnetic and electric dipole moments at the
$\tau \tau \gamma$ vertex. No interferences are neglected and no
fermion masses are set to zero.  The calculation shows that the
total cross section as well as the shapes of the photon energy and
angular distributions are sensitive to anomalous magnetic
and electric dipole moments.  We also find
that  interference between anomalous and Standard Model
contributions is  significant for the  case of the anomalous
magnetic dipole moment, and increases in  importance for smaller
values $F_2(0)$. The bulk of the contribution from terms linear
in $F_2(0)$ arises from interference between anomalous and Standard
Model final states.  Contributions from the electric dipole moment
evidently do not interfere. These results are in excellent
agreement with a simplified analytical calculation.

The calculation described here can be used to generate weights for 
different values of $F_2(0)$ and $F_3(0)$ on an event-by-event
basis. It is therefore an invaluable experimental tool, as it
facilitates proper simulation of detector effects and experimental
cuts in analysis of the process  
$e^+e^-\rightarrow\tau^+\tau^-\gamma$.  The matrix element
calculation is available from the  authors in the form of a FORTRAN
subroutine.

\begin{ack}
This work was supported by the National Science Foundation.
\end{ack}

\clearpage


\begin{thebibliography}{10}

\bibitem{COHEN87A}
E.~R. Cohen and B.~N. Taylor,
\newblock Rev. Mod. Phys. {\bf 59}, 1121 (1987).

\bibitem{ABDULLAH90A}
K.~Abdullah et~al.,
\newblock Phys. Rev. Lett. {\bf 65}, 2347 (1990).

\bibitem{BAILEY78A}
J.~Bailey et~al.,
\newblock Journ. Phys. {\bf G4}, 345 (1978).

\bibitem{SAMUEL91A}
G.~L. M.~A.~Samuel and R.~Mendel,
\newblock Phys. Rev. Lett. {\bf 67}, 668 (1991),
\newblock Erratum {\it ibid} {\bf 69},95 (1992).

\bibitem{SILVERMAN83A}
D.~J. Silverman and G.~L. Shaw,
\newblock Phys. Rev. {\bf D27}, 1196 (1983).

\bibitem{ESCRIBANO97A}
R.~Escribano and E.~Mass{\'{o}},
\newblock Phys. Lett. {\bf B395}, 369 (1997).

\bibitem{ESCRIBANO94A}
R.~Escribano and E.~Mass{\'{o}},
\newblock Nucl. Phys. {\bf B429}, 19 (1994).

\bibitem{ESCRIBANO93A}
R.~Escribano and E.~Mass{\'{o}},
\newblock Phys. Lett. {\bf B301}, 419 (1993).

\bibitem{GRIFOLS91A}
J.~A. Grifols and A.~M{\'{e}}ndez,
\newblock Phys. Lett. {\bf B255}, 611 (1991),
\newblock Erratum {\it ibid} {\bf B259}, 512 (1991).

\bibitem{DELAGUIA90A}
{F. del Aguia and M. Sher},
\newblock Phys. Lett. {\bf B252}, 116 (1990).

\bibitem{KINOSHITA81A}
T.~Kinoshita and W.~B. Lindquist,
\newblock Phys. Rev. Lett. {\bf 47}, 1573 (1981).

\bibitem{HUGHES85A}
V.~W. Hughes and T.~Kinoshita,
\newblock Comm. Nucl. and Part. Phys. {\bf 14}, 341 (1985).

\bibitem{FORM}
J.~Vermaseren,
\newblock FORM is a symbolic Manipulation Program by J. Vermaseren. The free
  version 1 can be obtained by anonymous ftp from {\tt{ftp.nikhef.nl}} in
  directory {\tt{pub/form}}. \\ See also: Symbolic Manipulation with FORM,
  version 2, ISBN 90-74116-01-9, (1991), CAN, (Computer Algebra Nederland),
  Kruislaan 419, 1098 VA, Amsterdam.

\bibitem{IMPROVED_BORN}
{M. Consoli and W. Hollik},
\newblock {Electroweak Radiative Corrections for Z Physics},
\newblock in {\em {Z Physics at LEP1, Vol. 1}}, edited by {G. Altarelli, R.
  Kleiss and C. Verzegnassi}, 1989,
\newblock CERN Report CERN-89-08.

\bibitem{FOWL}
F.~James,
\newblock FOWL and GENBOD,
\newblock CERN Program Library, Long writeups W\,505, (1972, revised 1981) and
  W\,515 (1975).

\bibitem{KORALZ}
{S. Jadach, B. F. L. Ward and Z. W{\c{a}}s},
\newblock Comp. Phys. Comm. {\bf 79}, 503 (1994).

\bibitem{SWAIN96A}
{S. S. Gau, T. Paul, J. Swain and L. Taylor},
\newblock ``${\mathrm{e^+e^-\rightarrow\tau^+\tau^-\gamma}}$ with anomalous
  couplings'', Invited talk at the L3 Plenary Meeting (presented by J. Swain),
  28-30 May 1996, DESY Zeuthen.

\bibitem{BIEBEL96A}
J.Biebel and T.~Riemann,
\newblock Z. Phys. {\bf C76}, 53 (1997).

\bibitem{TAU96_L3}
{L. Taylor (for the L3 Collaboration)},
\newblock {Measurement of the Tau Anomalous Magnetic Moment},
\newblock in {\em Proceedings of the TAU\,96 Workshop}, edited by {J. Smith and
  W. Toki},
\newblock Nucl. Phys. B (Poc. Suppl.) 55C, 285 (1997).

\bibitem{TAUOLA}
{S. Jadach, Z. W{\c{a}}s, R. Decker, J.H. K{\"{u}}hn},
\newblock Comp. Phys. Comm. {\bf 76}, 361 (1993).

\bibitem{L3_GEANT}
The L3 detector simulation program is based on GEANT Version 3.15 (see: R. Brun
  {\em et al.}, ``GEANT 3'', CERN DD/EE/84-1 revised, (1987)). GHEISHA is used
  to simulate hadronic interactions (see: H. Fesefeldt, RWTH Aachen Report
  PITHA 85/02 (1985)). This program allows for the effects of energy loss,
  multiple scattering, decays and interactions in the detector material, as
  well as for time-dependent detector effects.

\end{thebibliography}
\end{document}